\documentclass[preprint,showpacs,showkeys,preprintnumbers,amsmath,amssymb,superscriptaddress]{revtex4}

\usepackage{graphicx}
\usepackage{dcolumn}
\usepackage{bm}
\usepackage[latin1]{inputenc}

\begin{document}

\author{J. G. Oliveira Jr.}
\affiliation{Centro de Forma\c{c}\~{a}o de Professores, Universidade
Federal do Rec\^{o}ncavo da Bahia, 45.300-000, Amargosa, BA, Brazil}
\author{R. Rossi Jr.}
\affiliation{Departamento de F\'{\i}sica, Instituto de Ciências
Exatas, Universidade Federal de Minas Gerais, C.P. 702, 30161-970,
Belo Horizonte, MG, Brazil}
\author{M. C. Nemes}
\affiliation{Departamento de F\'{\i}sica, Instituto de Ciências
Exatas, Universidade Federal de Minas Gerais, C.P. 702, 30161-970,
Belo Horizonte, MG, Brazil}

\title{Protecting, Enhancing and Reviving Entanglement}

\begin{abstract}
We propose a strategies not only to protect but also to enhance and
revive the entanglement in a double Jaynes-Cummings model. We show
that such surprising features arises when Zeno-like measurements are
performed during the dynamical process.
\end{abstract}

\maketitle

Entanglement, ever since the conception of Quantum Mechanics by E.
Schrödinger, represents the most counterintuitive phenomenon
predicted by the theory. Nowadays, given the impressive advance in
technology physicists are able to observe entanglement in their
laboratories \cite{art1,art2,art3,art4} and also to show that it
represents a key ingredient for several important technological
applications. For example, the development of the quantum
information \cite{art5} which is intimately linked to the control
and manipulation of entanglement.

In a recently reported work Zeno-like measurements have been used to
protect an entangled state from its environment \cite{art6}. Usually
QZE is used as a tool for: Error prevention
\cite{art7,art8,art9,art10,art11}, quantum state engineering
\cite{art12}, state preservation \cite{art13} and decoherence
control \cite{art14,art15,art16}.

In the present contribution we show, in the context of double
Jaynes-Cummings model \cite{art17}, new results arising from the
combination of entanglement dynamics and Zeno-like measurements. The
results are strategies to control and enhance entanglement
\emph{even after sudden death} (induced revival).

We show that Zeno-like measurements, in the context of a double
Jaynes-Cummings model, may enhance the entanglement of the state
$\alpha|11\rangle + \beta|00\rangle$ ($\alpha >\beta$), bringing it
to a Bell state, even after time destruction of quantum correlations
(sudden death). The success of this procedure is conditioned to the
measurement result.

We have also shown that QZE can inhibit the entanglement dynamics in
duble Jaynes-Cummins system.

Let us consider a four qubits system coupled 2($ab$) by 2($AB$) with
identical coupling coefficient($g$). The interaction is such that it
couples $a$ only to $A$, and $b$ to $B$ as in Jaynes-Cummings models
quoted above.

Let us consider the initial state
$|\phi_{+}\rangle_{ab}|0,0\rangle_{AB}=(\alpha_{0}|1,1\rangle_{ab}+\beta_{0}|0,0\rangle_{ab})|0,0\rangle_{AB}$,
whose dynamical evolution is given as

\begin{eqnarray}
&&|\phi_{+}\rangle_{ab}|0,0\rangle_{AB}\longrightarrow\left(\alpha_{0}a^{2}(t)|1,1\rangle_{ab}+\beta_{0}|0,0\rangle_{ab}\right)|0,0\rangle_{AB}\label{evo2}
\\&&-\alpha_{0}b(t)\left[ia(t)(|1,0\rangle_{ab}|0,1\rangle_{AB}+|0,1\rangle_{ab}|1,0\rangle_{AB})+b(t)|0,0\rangle_{ab}|1,1\rangle_{AB}\right]\notag,
\end{eqnarray}
where $a(t)=\cos(gt)$ and $b(t)=\sin(gt)$. Note that at
$t=T=\frac{\pi}{2g}$ there is an entanglement swapping

\begin{equation}
|\phi_{+}\rangle_{ab}|0,0\rangle_{AB}\longrightarrow|0,0\rangle_{ab}|\phi_{-}\rangle_{AB},
\end{equation}
where
$|0,0\rangle_{ab}|\phi_{-}\rangle_{AB}=|0,0\rangle_{ab}(\alpha_{0}|1,1\rangle_{AB}-\beta_{0}|0,0\rangle_{AB})$.

Let us introduce a probe system composed of $N$ two level systems
all prepared on $|0\rangle$ state, i.e.
($\bigotimes_{k=1}^{N}|0^{(k)}_{M}\rangle$). This probe system
interacts one at the time (at times $t_{N}=N\tau$) with subsystem
$AB$. This interaction will discriminate, after each interaction,
between the two possibilities: no excitation in $AB$ and otherwise.
We also assume that the probe system can be measured after every
interaction with $AB$.

After a sequence of $N$ interactions keeping only those with null
result, i.e., the probe remains unexcited; the vector state will
become

\begin{equation}
|\Phi_{N}(T)\rangle=\frac{1}{\left(|\alpha_{0}|^2\cos^{4N}\left(g\tau\right)+|\beta_{0}|^2\right)^{1/2}}\left(\alpha_{0}\cos^{2N}\left(g\tau\right)|1,1\rangle_{ab}
+\beta_{0}|0,0\rangle_{ab}\right)|0,0\rangle_{AB},\label{evo3}
\end{equation}
where we fixed the total time of the evolutions as
$T=\frac{\pi}{2g}$ (the entanglement swap time), and
$\tau=\frac{T}{N}$.

It is apparent from this expression that the relative contribution
of the state $|1,1\rangle_{ab}$ can be manipulated. In what follows
we derive the conditions under which the entanglement enhancement of
this system is possible even after sudden death.

For this purpose we calculate the concurrency \cite{art18} of the
initial state

\begin{equation}
C_{0}=\frac{2|\alpha_{0}||\beta_{0}|}{|\alpha_{0}|^{2}+|\beta_{0}|^{2}}=2|\alpha_{0}||\beta_{0}|,\label{con0}
\end{equation}
and the one after $N$ measurements

\begin{equation}
C_{N}=\frac{2|\alpha_{0}||\beta_{0}|\cos^{2N}\left(g\tau\right)}{|\alpha_{0}|^{2}\cos^{4N}\left(g\tau\right)+|\beta_{0}|^{2}},\label{con1}
\end{equation}

Taking the limit $N\rightarrow\infty$ on (\ref{evo3}) and
(\ref{con1}), we can see the dynamics inhibition.

\begin{eqnarray}
\lim_{N\rightarrow\infty}|\Phi_{N}(T)\rangle&=&|\phi_{+}\rangle_{ab}|0,0\rangle_{AB},\\
\lim_{N\rightarrow\infty}C_{N}=C_{0}.
\end{eqnarray}
The vector state of the system after $N$ measurements tends to its
initial state as does the concurrence to its initial value when
$N\rightarrow\infty$. Therefore, in the Zeno limit the entanglement
dynamics on double Jaynes-Cummings model can be stopped, preserving
the initial entanglement in one of the parts. The calculation shows
that the initial entanglement contained in one of the partitions of
the system can be maintained as it is. Moreover it is also possible
to transfer this entanglement to the initially factorized partition
and preserve the entanglement there.

From equation (\ref{con0}) and the normalization of
$|\phi_{+}\rangle_{ab}|0,0\rangle_{AB}$
($\alpha_{0}|^{2}+|\beta_{0}|^{2}=1$), we may write:

\begin{eqnarray}
|\alpha_{0}|=
\begin{array}{c}
\sqrt{\frac{1}{2}+\sqrt{\frac{1-C_{0}^{2}}{4}}}; |\alpha_{0}|\geq|\beta_{0}|\\
\sqrt{\frac{1}{2}-\sqrt{\frac{1-C_{0}^{2}}{4}}}; |\alpha_{0}|\leq|\beta_{0}|\\
\end{array}\label{alpha}
\end{eqnarray}

Substituting (\ref{alpha}) in (\ref{con1}), and after some algebra
we may write the concurrence $C_{N}$ as function of the initial
concurrence and the number of measurements.

\begin{eqnarray}
C_{N}^{\pm}=
\begin{array}{c}
C_{N}^{+}; |\alpha_{0}|\geq|\beta_{0}|\\
C_{N}^{-}; |\alpha_{0}|\leq|\beta_{0}|\\
\end{array}
\end{eqnarray}
where

\begin{equation}
C_{N}^{\pm}=\frac{2C_{0}\cos^{2N}(g\tau)}{1+\cos^{4N}(g\tau)\mp\sqrt{1-C_{0}^{2}}[1-\cos^{4N}(g\tau)]}.
\end{equation}

The functions ($C_{N}^{+}$ and $C_{N}^{-}$) show a similar behavior
as $N\rightarrow\infty$, both tent to the initial concurrence value.
However, their behavior is quite different when $N$ is finite, this
is shown on Fig.1.

\begin{figure}[h]
\centering
  \includegraphics[scale=0.65,angle=-90]{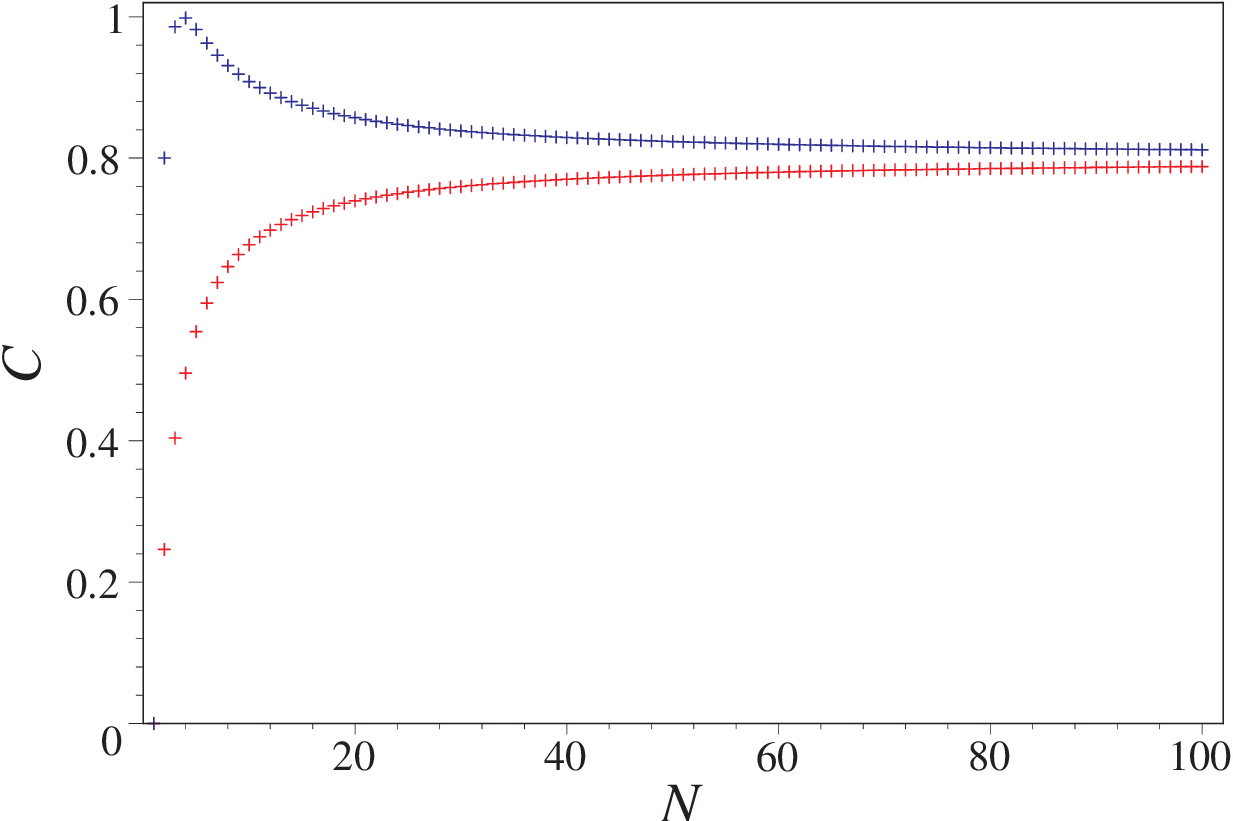}\\
  \caption{(Color online)Concurrence $C_{N}^{-}$ (lower curve) and $C_{N}^{+}$ (upper curve) as function of $N$ for $g\tau=\frac{\pi}{2N}$.}
\end{figure}

If only one measurement is performed at $\tau=\frac{\pi}{2g}$, no
inhibition can be observed and the entanglement is completely
transferred to the subsystem $AB$. Therefore, the concurrence of
subsystem $ab$ is null and both curves in Fig. 1 start from zero. As
$N$ increases $C_{N}^{-}$ grows steadily tending to $C_{0}$ when
$N\rightarrow\infty$. We conclude that for initial states with
$|\alpha_{0}|\leq|\beta_{0}|$, the effect of entanglement freezing
is the only one that can be induced.

On the other hand, the function $C_{N}^{+}$ shows an interesting
effect besides the entanglement freezing. When the initial state
respects the relation $|\alpha_{0}|>|\beta_{0}|$, the concurrence of
the subsystem $ab$ can exceed its initial value, and become very
close to the maximum value of 1. This is best visualized in the
coefficient $\alpha_{0}\cos^{2N}(g\tau)$ (in (\ref{evo3})) which
decreases as $N$ increases, and when its value approaches
$\beta_{0}$ the concurrence grows. If
$\alpha_{0}\cos^{2N}(g\tau)=\beta_{0}$ the concurrence is maximum
$C_{N}=1$.

It is also possible to bring the superposition $\alpha_{0}|11\rangle
+ \beta_{0}|00\rangle$ ($\alpha_{0} >\beta_{0}$) to a Bell state
with only one measurement. The procedure is simple, one just has to
let the system evolve freely and perform a measurement on subsystem
$AB$ at the time $t=\frac{1}{g}\arccos(\sqrt{\beta_{0}/\alpha_{0}})$
(when $\alpha_{0}a^{2}(t)=\beta_{0}$). Selecting the states with
null result, one will have a Bell state prepared in subsystem $ab$.
This procedure is conditioned to the measurement result and can be
performed even after the sudden death of entanglement as will be
shown.

The appearance of entanglement sudden death for the initial states
$(\alpha|11\rangle + \beta|00\rangle)|0,0\rangle$ ($\alpha
>\beta$), in the context of the double Jaynes-Cummings model is well know \cite{art9}. In order to explicitate this
effect we give the expression for the concurrence ($C_{f}$) of the
state evolving without external interference, as a function of the
initial concurrence.

\begin{equation}
C_{f}^{\pm}(t)=\max(0,\Lambda^{\pm}(t)),
\end{equation}

where

\begin{equation}
\Lambda^{\pm}(t)=\sqrt{1\pm\sqrt{1-C^{2}_{0}}}\cos^{2}(gt)\left(\sqrt{1\mp\sqrt{1-C^{2}_{0}}}-\sqrt{1\pm\sqrt{1-C^{2}_{0}}}\sin^{2}(gt)\right).
\end{equation}

Notice that for times

\begin{equation}
t=\frac{1}{g}\arcsin\left(\frac{\sqrt[4]{1-\sqrt{1-C^{2}_{0}}}}{\sqrt[4]{1+\sqrt{1-C^{2}_{0}}}}\right),
\end{equation}
which are prior to the swap time, there will be a sudden death of
entanglement. Fig.2 illustrate this. However, in Fig.3 we show how a
single measurement can induce an early entanglement resurrection.

The graphic on Fig.3 shows the concurrence ($C_{1}^{+}$) after a
single measurement on $AB$ with null result performed at times $t$,
for different values of the initial concurrence of the states
$(\alpha|11\rangle + \beta|00\rangle)|0,0\rangle$ ($\alpha
>\beta$). Notice that (from Fig.2) the concurrence is zero after the sudden death time. In Fig.3,
surprisingly enough, the set of vanishing final concurrencies are
``brought to life" by a single measurement performed not necessarily
immediately after the time at which sudden death takes place.

\begin{figure}[h]
\centering
  \includegraphics[scale=0.5]{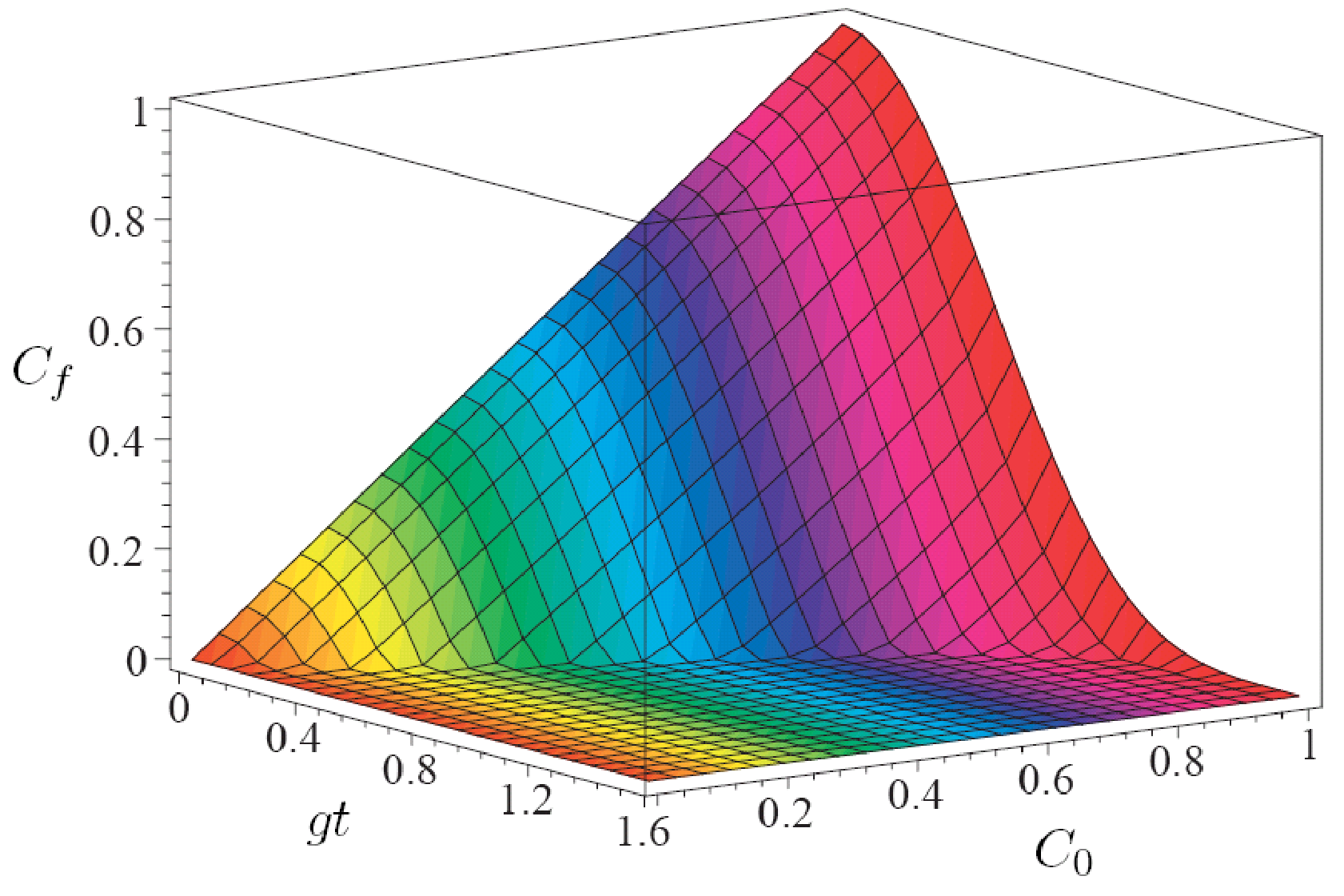}\\
  \caption{(Color online)Concurrence without measurements ($C_{f}$) as function of $gt$ and $C_{0}$ for initial states $(\alpha|11\rangle + \beta|00\rangle)|0,0\rangle$ ($\alpha
>\beta$) .}
\end{figure}

\begin{figure}[h]
\centering
  \includegraphics[scale=0.5]{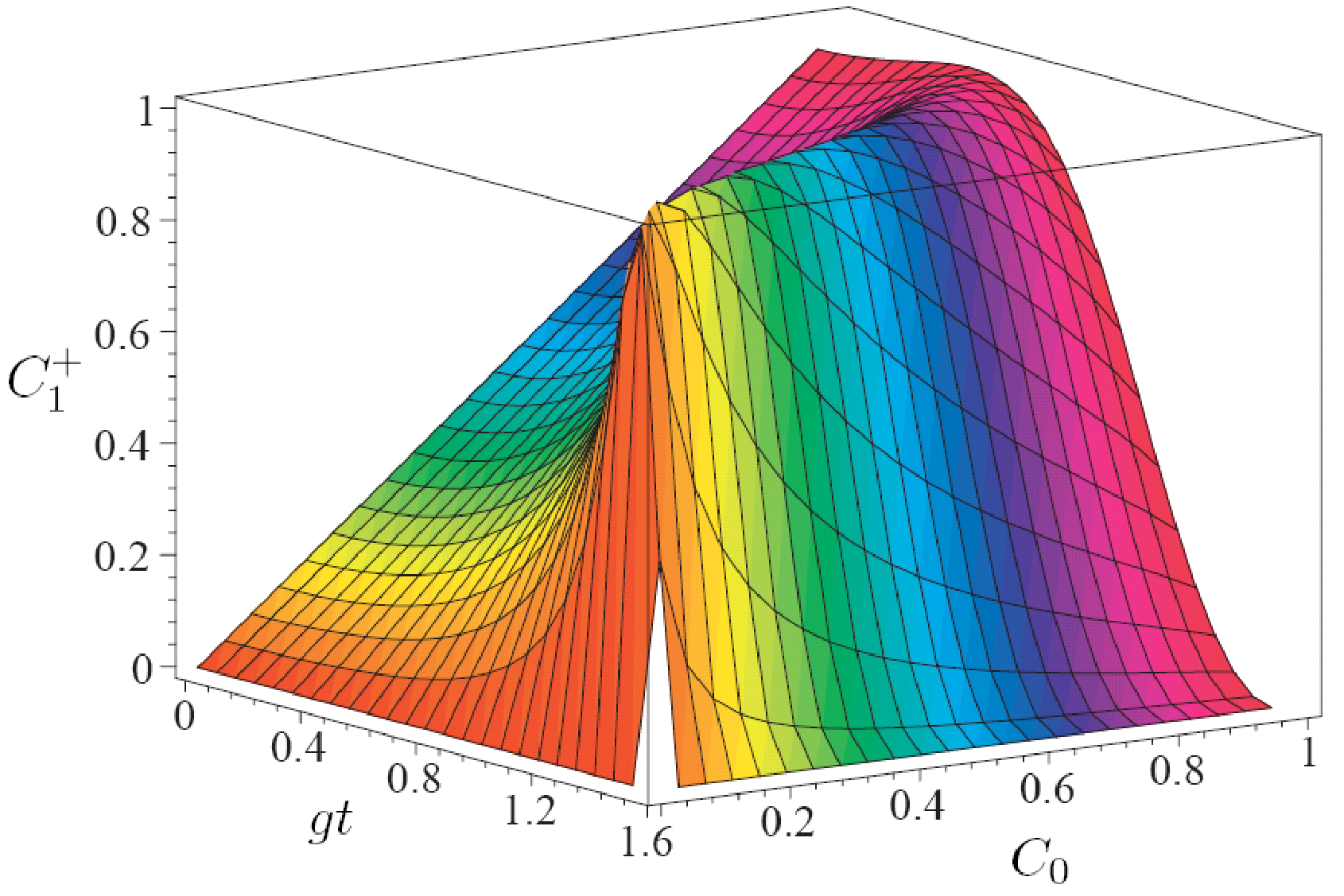}\\
  \caption{(Color online)Concurrence ($C_{1}^{+}$) after a
single measurement on $AB$ with null result, as function $gt$ and
$C_{0}$ for initial states $(\alpha|11\rangle +
\beta|00\rangle)|0,0\rangle$ ($\alpha
>\beta$) .}
\end{figure}

The reason for this is that after the sudden death and before the
entanglement swap time, there is no entanglement on $ab$, but the
excitations are not completely transferred to $AB$. Then if one
performs a measurement on $AB$, and gets a null result, a state with
finite entanglement is prepared in $ab$. It is even possible to
resurrect the entanglement to the maximum value ($C_{1}^{+}=1$),
exceeding the initial one, if the measurement is performed at
$t=\frac{1}{g}\arccos\left(\sqrt{\beta_{0}/\alpha_{0}}\right)$.

To summarize, we have shown that QZE can be used as a tool to
inhibit entanglement dynamics in the context of double
Jaynes-Cummings model. We also show that Zeno-like measurements are
capable of enhancing (conditionally) the entanglement of the state
$(\alpha|11\rangle + \beta|00\rangle)|0,0\rangle$ ($\alpha
>\beta$) and apply this procedure to avoid sudden death observed
in other calculations \cite{art19}. Entanglement freezing is
possible for all the others Bell states, however not the
enhancement.

The authors acknowledge financial support by CNPq.


\begin{thebibliography}{19}


\bibitem{art1} A. Aspect, J. Dalibard and G. Roger \prl \textbf{49}, 1804 (1982).


\bibitem{art2} A. Auffeves, P. Maioli, T. Meunier, S. Gleyzes, G. Nogues1, M. Brune, J. M. Raimond and S. Haroche,
\prl \textbf{91},  230405 (2003).


\bibitem{art3} A. Rauschenbeutel, P. Bertet, S. Osnaghi, G. Nogues, M. Brune, J. M. Raimond and S. Haroche,
 \pra \textbf{64}, 050301 (2001).


\bibitem{art4}  R. McDermott, R. W. Simmonds, M. Steffen, K. B. Cooper, K. Cicak,
K. D. Osborn, Seongshik Oh, D. P. Pappas, J. M. Martinis,
\textit{Science} \textbf{307}, 1299 (2005).



\bibitem{art5} M. A. Nielsen and  I. L. Chuang, {\it Quantum Computation
and Quantum Information} (Cambridge University Press, Cambridge,
England, 2000).



\bibitem{art6} S. Maniscalco, F. Francica, R. L. Zaffino, N. Lo Gullo and F. Plastina,
\prl \textbf{100}, 090503 (2008).



\bibitem{art7} Chui-Ping Yang and Shih-I Chu, \pra \textbf{66}, 034301 (2002).


\bibitem{art8} L. Vaidman, L. Goldenberg, and S. Wiesner, \pra \textbf{54},  R1745 (1996).


\bibitem{art9} Lu-Ming Duan and Guang-Can Guo, \pra \textbf{57}, 2399 (1998).


\bibitem{art10} WonYoung Hwang, Hyukjae Lee, Doyeol (David) Ahn, and Sung Woo Hwang, \prl \textbf{62}, 062305 (2000).


\bibitem{art11} N. Erez, Y. Aharonov, B. Reznik, and L. Vaidman, \pra \textbf{69}, 062315 (2004).


\bibitem{art12} M. Zhang and L. You, \prl \textbf{91}, 230404 (2003).


\bibitem{art13} D. Dhar, L. K. Grover, and S. M. Roy, \prl \textbf{96}, 100405 (2006).


\bibitem{art14} O. Hosten, M. T. Rakher, J. T. Barreiro, N. A. Peters and P. G. Kwiat,
Nature \textbf{439}, 949 (2006).


\bibitem{art15} J. D. Franson, B. C. Jacobs, and T. B. Pittman, \pra \textbf{70}, 062302 (2004).


\bibitem{art16} P. Facchi, S. Tasaki, S. Pascazio, H. Nakazato, A. Tokuse and D. A. Lidar,
\pra \textbf{71}, 022302 (2005).



\bibitem{art17} I. Sainz and  G. Björk, \pra, \textbf{76}, 042313
(2007).



\bibitem{art18} S. Hill and W. K. Wootters, \prl, \textbf{78}, 5022 (1997).


\bibitem{art19} M. Yönaç, T. Yu and  J. H. Eberly, J. Phys.
B, \textbf{39}, S621 (2006).



\end{thebibliography}
\end{document}